\documentstyle[11pt,newpasp,twoside,psfig]{article}
\def\mag{\hbox{$^{\rm m}$}}

\begin{document}

\title{Protoplanetary disks around Herbig Ae/Be stars: 
Indications from ISO spectroscopy}
\author{M.E. van den Ancker}
\affil{Harvard-Smithsonian Center for Astrophysics, 60 Garden Street, 
MS 42, Cambridge, MA 02138, USA}

\begin{abstract}
An analysis of solid-state features in infrared spectra of 
46 Herbig Ae/Be stars is presented. The presence of solid-state 
emission bands is compared to other indicators of 
circumstellar material, such 
as H$\alpha$ emission, optical variability and sub-mm continuum 
fluxes. The correlation between these different indicators is 
weak, if present at all, in our sample. However, a strong 
dependence on spectral type of the central star seems to 
be present: stars with spectral type earlier than B9 show either 
amorphous silicate in absorption or infrared spectra dominated 
by PAH emission, whereas more than 70\% of the stars of later 
spectral type show silicate emission. We conclude that the 
infrared spectrum of Herbig Be stars is in general dominated 
by emission from the circumstellar envelope, whereas the 
lower-mass Herbig Ae stars show a spectrum that is dominated 
by a disk that is passively heated by the central star.
\end{abstract}

\section{Introduction}
Herbig Ae/Be stars are young intermediate-mass (2--10~M$_\odot$) stars 
which are still surrounded by gas and dust from their natal cloud.  
Many possess circumstellar disks which are believed to 
be the site of on-going planet formation.  The dust in these 
circumstellar disks, heated by the central star and possibly by 
viscous heating of material that is being accreted, shows up 
as excess emission above photospheric levels at infrared to sub-mm 
wavelengths.  The circumstellar gas can be traced in spectral 
lines, of which H$\alpha$ is most prominent.  Large-amplitude ($>$ 1\mag) 
variations in optical brightness are seen in some Herbig Ae stars, 
and are commonly ascribed to circumstellar dust clouds moving in and 
out of our line of sight towards the central star.
Although the optical to sub-mm energy distribution of Herbig Ae/Be stars 
has been well explored by previous authors (e.g. Hillenbrand et al. 1992), 
the chemical and mineralogical composition of the dust remained 
poorly studied until the 1995 launch of the {\it Infrared Space
Observatory} (ISO; Kessler et al. 1996).  This first possibility to 
study the complete infrared spectrum of these objects in detail revealed 
a large variety in dust properties, from small aromatic hydrocarbons 
to silicate dust.  Moreover, some sources were shown to contain 
partially crystalline dust grains, similar to those found in 
comets in our own solar system (Malfait et al. 1998, 1999; van den 
Ancker et al. 1999; Meeus et al., these proceedings). 
Here, the first inventory of solid-state features in all 
Herbig stars observed by the {\it Infrared Space Observatory} 
is presented and we investigate their correlation with more traditional 
tracers of circumstellar material.
\begin{figure*}[t]
\centerline{\psfig{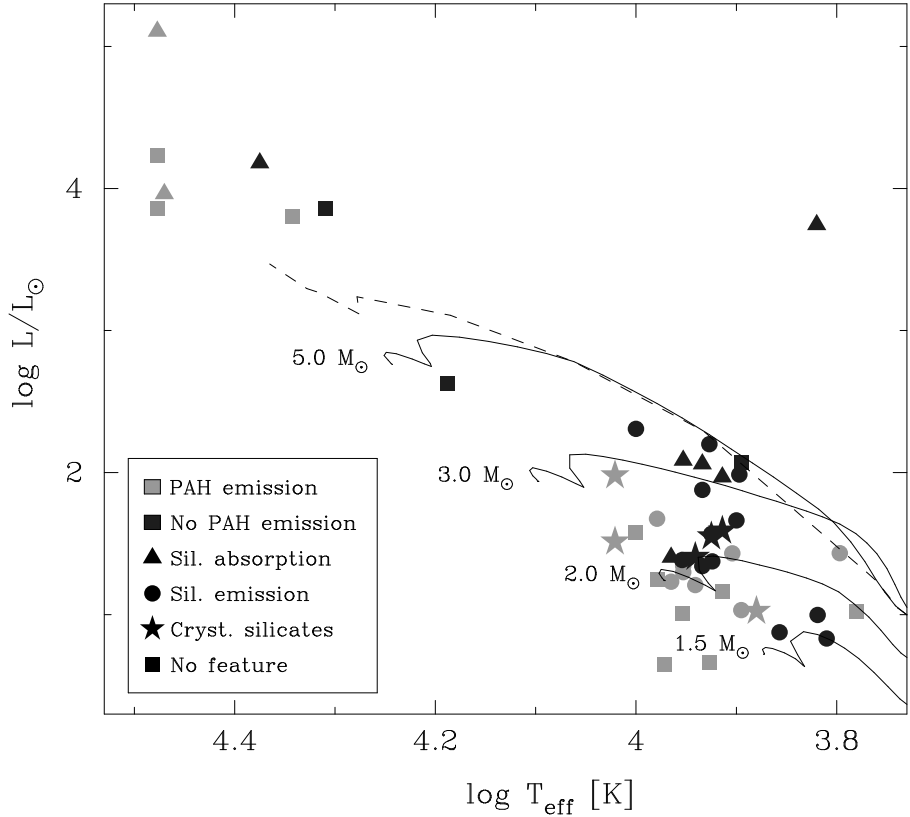}}
\caption[]{HRD of the stars in our sample. 
Plot symbols indicate the solid state components
present in the ISO spectra (see caption).
Also shown in the figure are the pre-main
sequence evolutionary tracks (solid lines) and the birthline 
(dashed line) by Bernasconi (1996).}
\end{figure*}

\section{Data Analysis}
An inspection of the ISO data archive revealed the presence
of spectroscopic data on 46 Herbig Ae/Be stars, obtained with 
the short-wavelength spectrometer (SWS) and the 
spectroscopic mode of the photometer (ISOPHOT). 
Spectra were retrieved and reduced, after 
which they were inspected for the following features: (a)
the emission bands at 3.3, 3.4, 6.2, 7.6, 7.8, 8.6, 11.3 and
12.7~$\mu$m, often attributed to polycyclic aromatic hydrocarbons (PAHs), 
(b) the broad band around 10~$\mu$m due to amorphous silicates, 
and (c) sharper emission bands at 10.2, 11.4, 16.5, 19.8, 23.8, 27.9 
and 33.7~$\mu$m due to crystalline silicates.
Using these data, we investigated the 
correlation of infrared spectral features with parameters of the 
systems from literature ($T_{\star}$, $L_{\star}$, level of optical 
variability, H$\alpha$ profile, and dust masses as traced by sub-mm fluxes).

\section{Discussion and conclusions}
The strongest correlation found is between spectral type of
the central star and silicate emission: Herbig stars of spectral
type earlier than B9 show silicate absorption, whereas a large
majority ($\approx$70\%) of the Herbig Ae stars of later type
show silicate emission.  Since strong optical variability
due to variable circumstellar extinction is also only found in
Herbig stars with spectral type of B9 or later (van den Ancker 
et al. 1998), both phenomena
may be related.  However, no correlation between the level of
optical variability and silicate emission could be found,
perhaps due to our limited sample size.

In most sources with absorption due to amorphous silicates, 
we also observe the absorption bands due to H$_2$O and CO$_2$ 
ice, with a relative strength comparable to that in the 
interstellar medium. However, two sources (Z~CMa and V645~Cyg) 
show strong silicate absorption, but no evidence for water or 
CO$_2$ ice bands, demonstrating the chemical evolution that 
has taken place in the circumstellar environment of 
these objects.

No strong correlation between spectral type and PAH emission
could be found. This is surprising, since the excitation of
PAH molecules is thought to require intense ultraviolet radiation
fields.  This means that on average the particles responsible
for the PAH emission must be closer to the central star, and
hence suffer less geometric dilution of the stellar radiation
field, in Herbig Ae stars than in Herbig Be stars.

Both the differences in silicate and in PAH behaviour can be
explained by assuming that the infrared spectrum of Herbig Be
stars is in general dominated by their circumstellar envelope
rather than a disk. In contrast, the more slowly evolving
Herbig Ae stars have time to disrupt their envelope and their
spectrum may be dominated by thermal emission from the 
protoplanetary disk.

Crystalline silicates, as are also found in comets in our own
solar system, are visible in 15\% of the late-type Herbig stars,
all of systems that are relatively isolated and appear to be
relatively old (a few million years). Therefore also in young
stars longevity appears to be a prerequisite for the annealing
process. These systems form a close analog to the young solar
system and may provide the strongest clue to date that the same
processes that have led to rocky planets in our own solar system
are also taking place around other stars.

\end{document}